# BIG TECH COMPANIES' IMPACT ON RESEARCH AT THE FACULTY OF INFORMATION TECHNOLOGY AND ELECTRICAL ENGINEERING

**Research Ethics Course Report**

Submitted By
**Group 2**

Submitted On
**12/10/2021**

**Group 2 Members**
Ahmad Hassanpour, An Thi Nguyen, Anshul Rani, Sarang Shaikh, Ying Xu, Haoyu Zhang

# 1. Introduction

Artificial intelligence is gaining momentum, ongoing pandemic is fuel to that with more opportunities in every sector specially in health and education sector. But with the growth in technology, challenges associated with ethics also grow (Katharine Schwab, 2021). Whenever a new AI product is developed, companies publicize that their systems are transparent, fair, and are in accordance with the existing laws and regulations as the methods and procedures followed by a big tech company for ensuring AI ethics, not only affect the trust and perception of public, but it also challenges the capabilities of the companies towards business strategies in different regions, and the kind of brains it can attract for their projects. AI Big Tech companies have influence over AI ethics as many influencing ethical-AI researchers have roots in Big Tech or its associated labs such as one led by Gebru and Mitchell (Gofman, Michael, and Zhao Jin, 2020).

Big tech influence on AI ethics and thus dominance on academic research starts with attracting brilliant minds for direct hiring, luring academic researchers with great funding, and having in-house computational resources (Taebi et al., 2019). According to the report published by HAI (https://hai.stanford.edu/research/ai-index-2021) 65% of North American AI PhDs now go to industries, which is a 21% increase from the figure reported in 2010 (https://hai.stanford.edu/research/ai-index-2021). This numeric data shows the potential influence of big tech companies in both AI research and implementation. Recently, Timnit Gebru, a renowned researcher in the AI world and co-lead of AI ethics team of Google was made to quit her job in December 2020 (Gebru, 2021), as Gebru got into conflicts with an organization for a coauthored research paper (Bender, 2021). This is an example of direct dominance of firms. This paper described risks associated with AI models and defined potential biases. This paper examined the risks of the AI models which are used involved in google search queries. Paper outlined the existing potential biases in models based on the race, gender, westernization and more. Google insisted her to retract the paper, or the author should not have Google affiliation. Instead of going for an open discussion with Gebru as asked by her, google announced her resignation. Later, it was said that her work was not up to the mark despite her great history in the same field.

Despite the existence of dominance and influence of big tech companies, scientific research is supposed to be supportive and in alliance with the future needs of society. An increase in government policies related to more industry-academia collaborations in many countries is the result of this notion. In many cases, the prospect of universities and individual researchers to pursue research activities depends upon such collaborations. In result, such policies lead to dominance of big tech companies. The challenge is that not many alternatives are available for AI research, as AI project requires huge funding for involved high processing resources. Thus, AI projects are led by limited companies (Big Tech) who can afford it. Now, a question arises of how the collaboration guidelines and procedures between industry and academia should be framed. This report investigates the same question: Section 2 details how big tech companies dominate universities and individual researchers, and eventually manipulate the AI ethics field. Section 3 presents the recommendations and measures to mitigate/avoid this dominance.

## 2. Impacts of Big Tech dominance

### 2.1 Big Tech is impacting academic freedom

Traditionally, academic freedom has meant that scholars in academic institutions should have their own freedom to pursue the knowledge they want in their fields. While Ashford (Ashford, 1983) had a modern definition of academic freedom which involved concern to both university and society. He categorized academic freedom into four types in science and technology: intellectual inquiries conducted within existing paradigms, intellectual inquiries that challenge existing paradigms, inquiries directed primarily at the socioeconomic or political consequences of science and technology, and inquiries directed at producing foreseeable commercial value. The latter two would be categorized as applied research. We would like to discuss what happens when academic institutions are cooperating with Big Tech.

Big Tech has been working hard in building connections with leaders in academia. The Vector Institute of the University of Toronto had heads of AI branches of various companies in their faculty, including Google, Uber, Nvidia, and other Big Tech companies, which led the directions of AI research. The academic institutions would remain quiet regarding any ethical issues caused by Big Tech companies that provided the fundings (Abdalla and Abdalla, 2021) despite two-thirds of their fundings being from the governments (The Canadian Press, 2017; Lee, 1996). Lee (1996) also argued that losing freedom in academic research equals losing pursuing long-term, disinterested and fundamental research.

There is no doubt that Big Tech companies sponsor academic institutions for commercial reasons. These AI-leading companies want to gain benefit from the collaboration with universities. At the same time, universities strive for knowledge generation and dissemination for a better society. The research direction must be tailored to increasing the funding companies' interest in order to fulfill the requests on profit from Big Tech. Therefore, academic institutions have no choice but to lose some part of their freedom because of the cooperation with industrial companies (Hillerbrand and Werker, 2019).

For individual researchers, it is possible to have non-disclosure agreements with funding companies when cooperating with them. This kind of agreement would always align with the business interests of the companies and protect their benefits. Although the internal commitment does not indicate that researchers could not act with ethics or integrity, the Big Tech has more power to decide which good things to do for their project. According to studies (Karlsson et al., 2019; Google, 2021; Hagendorff, 2020), Big AI companies are rarely controlled by published AI ethics principles and guidelines. In fact, there are no complete and suitable laws for dangerous AI products in the markets and society nowadays. Individual researchers working for the industrial cooperation project often feel powerless when there are conflicts between Big Tech and the institution. These contradictory goals are mainly about balancing the profits and publications.

In summary, AI Big Tech companies are impacting academic freedom to a certain extent on both academic institution and individual researchers. It is problematic that a publicly funded research institution fails to speak out on ethical issues, which is against the public interest.

## 2.2 Big Tech is stealing top researchers from academia

Colleges exist, partially, to meet the requirements of industry, but the request for talented AI analysts is surpassing the supply. In reality, requests at companies such as Google, Facebook, and Amazon seem to take off less gifted researchers to educate the following era, driving to academics' fear. Scores of skilled researchers have cleared out or passed up college posts for pay rates two to five times higher at major innovation firms, where other than getting way better pay, unused initiates can take on real-world issues with computer control and datasets that the scholarly world cannot trust to supply. The effect of the contracting craze is uncovered in a confidential Guardian survey overview of the UK's first-class Russell Bunch colleges, which found that numerous best educate were battling to keep up with the request from tech firms that are forcefully extending their AI inquire about research groups (UniversityWorldNews, 2017).

## 2.3 Big Tech is influencing the type of academic AI research that is pursued

Abdalla and Abdalla (2021) argue that Big Tech companies can influence the type of academic AI research that is pursued through the funding they provide to academic researchers. In Google's leaked documents, their researchers are told to "strike a positive tone". Furthermore, Google's researchers are required to consult with policy and public relations teams prior to publishing their research. While there is no evidence that academic researchers who receive funding from Big Tech companies are bound to the same guidelines, it is not unlikely that analogous public relations assessments are made when these companies approve grants to academic researchers (Abdalla and Abdalla, 2021).

Another way that Big Tech can influence which type of academic research is pursued, is through computational infrastructures. By providing clouds, mobile devices, chips and application programming interfaces (APIs) as the basis for all AI-based software development, power has been concentrated at the hands of a few major corporations with infrastructures and processing power. Not only is this invasive in terms of processing of individuals' personal data, but it also necessarily requires that all software development must go through these major tech companies. Most academic researchers do not have access to the large-scale data and hardware infrastructure, and they therefore have to use the provided infrastructures by Big Tech. For instance, the deep learning library called TensorFlow has been open-sourced by Google in 2016. For another example, Google-Colab which provides free hardware infrastructures has been introduced in 2018, again by Google. Furthermore, as Big Tech is sitting on massive amounts of data, which is needed to train large AI models, many academic AI researchers are dependent on industry collaboration. This allows Big Tech to decide what kind of research they will conduct.

The Big Tech influence on AI research is also extending to the AI ethics field. Abdalla and Abdalla (2021) estimate that 58% of AI ethics faculty in some of the top PhD granting universities (Massachusetts Institute of Technology, University of Toronto, Stanford, and Berkeley) have received funding from Big Tech. The authors further argue that Big Tech can influence what AI ethics researchers work on, as university faculty must cater their research to the views of Big Tech to receive funding (Abdalla and Abdalla, 2021). This is problematic as the AI ethics field should be able to criticize the underlying views that the technology of Big Tech is built on.

There does not necessarily need to be an intention from Big Tech to control the direction of AI ethics research for this shift to happen. As Abdalla and Abdalla (2021) argue, it is sufficient that "those applying for awards and those deciding who deserve funding do not share the same underlying views of what ethics is or how it 'should be solved'" (Abdalla and Abdalla, 2021). This pertains for instance to views on regulation, some AI ethics researchers arguing that "regulation may do little more than slow down the damage to society" (Arogyaswamy, 2020). Instead, they recommend that "Firms need to establish ethical standards" and argue that the users themselves are best equipped to protect their own rights (Arogyaswamy, 2020). These arguments are aligned with Big Tech's interests in avoiding regulation and shifting responsibility to the users.

Big Tech's influence on AI research can also happen through value drift in faculty members. Ashford (1983) argues that a faculty member who has worked on multiple industry-funded research projects might develop an "industrial viewpoint". This, in combination with Big Tech selecting which type of research to fund, can affect how research questions, discussions and conclusions are framed. Abdalla and Abdalla (2021) argue that a large portion of the research on fair AI is based on the Big Tech notion that ethical issues can be solved with technical research. For instance, a large focus in the field of AI fairness has been on developing mathematical definitions and tools to measure how fair an AI system is (Dwork et al., 2012; Hardt et al., 2016; Bellamy et al., 2019). However, some argue that solving the current issues in AI fairness requires human thought to fully understand the underlying social problems (Fazelpour and Lipton, 2020). This view has been in the minority in the AI fairness field.

## 2.4 Big Tech is controlling the academic discussion

By influencing the type of academic AI research that is pursued, Big Tech is also able to control the academic discussion. Abdalla and Abdalla (2021) argue that Big Tech's funding of major AI conferences and workshops also serves to shift the academic discussions in their favor. Out of the major workshops on AI ethics or fairness, only one of these does not have at least one organizer connected to Big Tech. As organizers they have the power to set the agenda of these conferences, thereby allowing Big Tech to control which aspects of ethics will be focused on (Abdalla and Abdalla, 2021). Much work has for instance been done on building fairer risk assessment tools for use in courts and the police by applying different fairness metrics or pre/post-processing the data (Coston et al., 2020; Mishler et al., 2021). However, discussions on whether predictive tools should be used to make decisions about human lives have generally been absent from these conferences.

## 2.5 AI ethics serves to legitimize Big Tech

The total effect of the above points is that much of the work in AI ethics serves to legitimize the activities of Big Tech, some of them pertaining to controversial technologies. As reported by Williams (2019), the European Commission's High Level Expert Group on AI, put together to give advice on ensuring "Trustworthy AI", consisted of 50 % industry representatives. DigitalEurope was one of these industry representatives, an organization boasting corporate members from Big Tech. When some experts proposed that technology such as autonomous lethal weapons and social credit score systems should be "red lines" for AI in Europe, some representatives wanted to formulate it differently. These topics are now instead labeled "critical concerns" (Williams, 2019).

Abdalla and Abdalla (2021) argue that Big Tech's support of various AI ethics initiatives, from symposiums and committees to guidelines for ethical AI, serves to reinvent their public image as socially responsible while allowing them to bypass legislation. Big Tech has announced its adherence to various AI ethics frameworks. These include Microsoft's Responsible AI principles, Google's AI principles, and Facebook's five pillars of Responsible AI (Microsoft, no date; Google, no date; Facebook, no date). The various ethics initiatives are not legally binding and an attempt to self-regulate to avoid actual legislation. In addition, there is evidence that Big Tech is actively searching for pro-industry academics who can be leveraged to battle legislation (Abdalla and Abdalla, 2021).

For another example, a group created by Big Tech (Microsoft, Google, Facebook, IBM, and Amazon) called Partnership on AI (PAI), despite characterizing itself as a "multistakeholder body" and claiming it is not a campaigning organization, hearing at the U.S. House Committee on Oversight and Government Change, the Partnership's official chief claimed that the organization is simply an asset to policymakers by conducting investigations that illuminate AI technologies and investigating the societal results of certain AI frameworks, as well as arrangements around the advancement and utilize of AI systems. It has been revealed that they directly affect the votes of individual elected officials. In November 2018, the PIA staff inquired academic researchers to contribute to a collective articulation to the Legal Committee of California with respect to a Senate charge on corrective change. The bill, within the course of killing cash safeguard, extended the utilize of algorithmic hazard appraisal in pretrial decision making, and required the Legal Committee to "address the recognizable. proof and moderation of any verifiable inclination in evaluation instruments. "The Association staff wrote, "we believe there's room to affect this legislation" (The intercept press, 2021).

## 3. Recommendations

In this section, we'll propose several recommendations for individual researchers, institutes, and the society regarding to the aspects of challenges mentioned in Section 2.

From the EU Charter & Code: 1.1 Research/Academic freedom is, "Researchers should focus their research for the good of mankind and for expanding the frontiers of scientific knowledge, while enjoying the freedom of thought and expression, and the freedom to identify methods by which problems are solved, according to recognised ethical principles and practices". But sometimes, researchers face some of the limitations to their academic freedom arising due to variety of factors like supervision, guidance, management, or budgetary reasons. In this circumstance, recommendations for maintaining academic freedom are:

- Individual academic researchers need to have full freedom regarding the research method selection, implementation, and dissemination of the results.
- The researchers must not have any pressure or demand from funding authorities (i.e., big tech companies).
- The author in (Bryden, 2013), concluded that it is better to have researchers the above freedom because only then they can enjoy doing research work freely as being as an autonomous self-regulatory research organization.
- The research involving political and administrative decisions must be performed in a professional and quality standard way maintaining high standards.
- The results of research must not be changed, tampered, or misused in any way to maintain the academic freedom of the researchers.

The brain drain of top researchers is challenging to solve as they can create massive profit for the Big Techs and correspondingly can be offered with considerable salary and research environment. The recommendations could be from different perspectives:

- The government should formulate regulations against the monopoly of Big Techs and restrict their business mode based on unethical usage of AI and information technology.
- Increase the investment on public investment on education and research to retain the researchers in academia. Specifically for AI research, more data and computational resources should be managed by public institution and reduce the dependency of research environment on Big Techs.
- Institute and individual researchers should take research ethics as a crucial part of their academic training.

Regarding to the influence from Big Tech on the type of AI research perused by academia, here are the proposed recommendations:

- As proposed in (Norwegian Association of Researchers, 2017) structural changes and reorganisation across the entire research sector and intentional deviation of research topics for competing the same research funds should be regulated to avoid narrowing the difference between HE sectors and the institutes and improve the diversity of research fields. This is especially crucial to reduce the influence of big tech on academic AI research, because as we shown in section 2 their fundings are usually focusing on specific fields based on their own profit instead of general social good and sustainability of the academic research.
- It will be particularly important to promote transparency in research and freedom of expression for researchers. This will ensure the possibility of supervision on ethical issue of AI research to prevent the misguiding based on common interest.
- Developing buyer expertise among external funders will also be critical so that they have more knowledge about what they can and cannot demand from the researchers. It is also argued in (EDRi and Access Now, 2020) that using only soft law approaches, voluntary certification schemes, and self-regulation by companies is not enough to mitigate the risks and laying down red lines and strict obligations by political response to the big techs.
- Specifically, AI research is highly dependent on the resources of data and computational power, which is usually monopolized by the big tech. Several regularizations have been issued to enhance privacy protection and restrict the big companies (e.g., GDPR (Voigt, 2017)), however, it is also essential to ensure these resources for the academic research to reduce the dependency of AI research on big techs and correspondingly reduce their influence.

To prevent of control of Big Techs on academic discussion:

- It is important to retain ownership of research results to the research institutes and researchers to prevent abusing by big techs without ethical responsibility.
- Similar as previous recommendation for research freedom, conferences and journals should also ensure the freedom and variation of research works, instead of flocking the topics interested by big tech for funding.
- Researchers should be free for the selection of relevant publication channels (i.e. journals, conference, etc) for the dissemination of the research results. Big tech companies must not influence/control the publication venues in order to justify their commercial/business needs. As argued in (Abdalla and Abdalla, 2021), "Statement
- Regarding Sponsorship and Financial Support" published at these workshops is not sufficient to avoid unconscious bias. A more general code for academic dissemination channel should be drafted or further improved to regulate the funding from big tech and evaluate the justification of the dissemination channel to prevent their manipulation on academic discussion.

Towards the legitimization of Big Techs based on AI researches, it is mainly caused by the common interest between industry and academia. Here are our suggestions:

- Improve the transparency of researchers and research works. Keep acknowledgment on the conflict of interest.
- Be aware of the bias of dissemination and discussion. The ethical issue of AI researches should not only be discussed by practitioners but also people with different background since it has shown influence on the whole society.

Finally, here are some general recommendations for researchers and research institutes about the ethical challenge under the dominance of Big Techs on AI research.

For individual researchers:

- Acknowledge the conflict of interest to prevent misguiding
- Avoiding biased experiments for satisfying the profit of your sponsor
- Include research questions about the impact of your research work on the society
- Involving actively to ethical discussions within other discipline and with people who have different background

As for research institutions:

- In the perspective of research institute, one of the most important rules is to let independent associations or institutes to supervise the research activities instead of sponsors with interest in the outcomes. (van Wee, 2019)
- Besides formalizing supervision policy and ethical codes, the institution should also attach significance to continuously education of research ethics as a crucial part of the academic training for researchers and students.
- Meanwhile a platform should be built to allow both internal and external individuals to share their opinion or have discussion on the ethical issues of research topics.
- Bias of representation from academia, industry and society should also be aware to avoid any dominant impact. Multidisciplinary discussion on the research ethics will also be beneficial to avoid common interest between the researcher and the sponsors within the same field and may also provide more various points of view.

## 4. Conclusion

In this report, we have investigated and illustrated the dominance of Big Tech companies on AI research and the impact within various aspects. It is shown that this dominance has led to severe challenges and potential risk to the society. Correspondingly, several recommendations are given to solve these problems.

Considering all the above recommendations and their strict use will help academic institutions and individual researchers in AI to reduce the impact or influence of funding authorities (I.e., big tech companies). More specifically, the researchers or academic institutes in AI will be able to get below things:

- More time of quality research and development (R&D) work, which will contribute towards pursuit of good knowledge
- Willingness of participation in public events and research platforms, which will help the on-going research to be more concrete
- Will create an environment for research-based teaching in academia
- Will prevent from intellectual dishonesty,
- Will preserve professional responsibility and prioritizations for quality research development in AI.

# 5. Group member contributions

In total, we had six group meetings to discuss about various aspects of the problem and assigned a part of the report to each person. The main contributions of each person have been mentioned in the below table. All of us went through the other sections and exchanged feedback to ensure we are covering the key issues of the problem.

| Name | Contribution |
|------|-------------|
| Anshul | Section 1 |
| Ying | Section 2.1 |
| An | Section 2.3-2.5 |
| Ahmad | Section 2.2, paragraph 2 of 2.3, and 2.5 |
| Haoyu | Section 3 |
| Sarang | Section 3 |
| All Group Members | Reviewed the complete report and were involved in discussions throughout the work. Equally contributed. |